\documentclass{aa}

\usepackage{graphicx}
\usepackage{txfonts}
\usepackage{amsfonts}
\usepackage{mathrsfs}
\usepackage{bm}
\usepackage{lipsum}
\usepackage{physics}
\usepackage{multirow}
\usepackage{rotating}
\usepackage{soul}
\usepackage{color}
\usepackage{ulem}

\def\bmv{\bm{\vup}}
\def\bmR{\bm R}
\def\bmnobs{{\bm n}_{\rm obs}}
\def\bhm{M_{\bullet}}

\def\civ{C\,{\sc iv}}

\def\bmn{\bm{n}_{\rm obs}}
\def\calA{{\cal A}}
\def\calB{{\cal B}}
\def\calC{{\cal C}}

\def\calO{{\cal O}}

\def\ccyan{\color{black}}

\def\ergs{{\rm erg\,s^{-1}}}
\def\fline{\ell_{\!\lambda}}
\def\Feii{Fe\,{\sc ii}}
\def\Fline{F_{\ell}}
\def\kms{\rm km\,s^{-1}}
\def\mathdotM{\dot{\mathscr{M}}}
\def\nobs{\bm{n}_{\rm obs}}

\def\Rg{R_{\rm g}}

\def\sunm{M_{\odot}}
\def\vup{\upsilon}

\begin{document}

\title{Spiral Arms in Broad-line Regions of Active Galactic Nuclei}

\subtitle{I. Reverberation and Differential Interferometric Signals of Tightly Wound Cases}

\author{Jian-Min Wang
    \inst{1, 2, 3}
    \and
    Pu Du
    \inst{1}
    \and
    Yu-Yang Songsheng
    \inst{1}
    \and
    Yan-Rong Li
    \inst{1}}

\institute{
    Key Laboratory for Particle Astrophysics,
    Institute of High Energy Physics, Chinese Academy of Sciences,
    19B Yuquan Road, Beijing 100049, China \\
    \and
    University of Chinese Academy of Sciences,
    19A Yuquan Road, Beijing 100049, China \\
    \and
    National Astronomical Observatories of China,
    Chinese Academy of Sciences, 20A Datun Road, Beijing 100020, China}

\abstract{
As a major feature in spectra of active galactic nuclei, broad emission lines
deliver information of kinematics and spatial distributions of ionized
gas surrounding the central supermassive black holes (SMBHs), that is the so-called
broad-line regions (BLRs). There is growing evidence for appearance of spiral arms
in the BLRs. It has been shown by reverberation
mapping (RM) campaigns that the characterized radius of BLRs overlaps with that
of self-gravitating regions of accretion disks. In the framework of the 
WKB approximation, we show robust properties of observational 
features of the spiral arms. 
The resulting spiral arms lead to various profiles of the broad emission line.
We calculate RM and
differential interferometric features of BLRs with $m=1$ mode spiral arms.
These features can be detected with high-quality RM and differential interferometric
observations via such as GRAVITY onboard Very Large Telescope Interferometer.
The WKB approximation will be
relaxed and universalized in the future to explore more general cases of density
wave signals in RM campaigns and differential spectroastrometry observations.
}


\maketitle

\section{Introduction}

Active galactic nuclei (AGNs) discovered by \cite{Seyfert1943} are
characterized by the appearance of prominent broad emission lines in their spectra
(e.g., see the composite spectra in \citealt{Berk2001} and \citealt{Hu2012}),
which usually have full widths at half maximum (FWHM) of $\gtrsim 10^{3}\,\kms$
or even a few $10^{4}\,\kms$ \citep[e.g., see reviews of][]{Ho2008,Netzer2013}.
Such broad widths of emission lines undoubtedly indicate very deep potential wells
in centers of AGNs first realized by \cite{Woltjer1959}, which directly motivated the
establishment of the scenario that accretion onto supermassive black holes
(SMBHs) powers the huge radiation energy of AGNs
\citep{ZelDovich1965,Salpeter1964,Lynden-Bell1969}. It is generally accepted
that broad emission lines stem from  BLR gas photoionzied by ionzing photons
emitted from accretion disks \citep{Osterbrock2006}. It has been made
great progress in understanding AGN and quasar activities \citep[e.g.,][]{Netzer2013}, 
but two major issues as to BLRs remain under debate so far: 1) origin of BLR gas; 
2) structure and dynamics \citep[see a brief review in][]{Wang2017}. 

Thanks are given to reverberation mapping (RM) campaigns since 1980s, with the
underlying principle proposed by \cite{Bahcall1972} and \cite{Blandford1982}.
Photons of emission lines from structured ionized gas undergo different paths to
observers, leading to time lags ($\tau$) of the emission lines with
respect to the ionizing photons. RM campaigns focusing on broad Balmer lines
had detected the anticipated lags in a number of Seyfert galaxies 
\citep[e.g.,][]{Peterson1993,Peterson1998, Bentz2013, Barth2015,U2021} and 
quasars \citep[e.g.,][]{Kaspi2000, Du2014, Du2018,
Shen2019} in the past decades. Along with the growing investment of
observing resources and the development of analytical algorithms, the general
geometry and kinematics of BLRs in some AGNs have been revealed by
velocity-resolved delay analysis \citep[e.g.,][]{Bentz2010, Denney2010, Grier2012, 
Grier2013,Du2016, U2021}, velocity-delay maps \citep[e.g.,][]{Xiao2018,Horne2021}, 
or dynamical modeling \citep[e.g.,][]{Bottorff1997,Pancoast2014,Li2018,Williams2020}. 
Many resolved BLRs have disk-like geometry of H$\beta$ line region (the 
other have inflow or outflow, or a kind of mixture of the three 
configurations)\footnote{It turns out that high-ionization lines, such as \civ\, 
line, favor origination from outflows. See \cite{Bottorff1997} for a first detailed 
study of \civ\, line in NGC 5548 observed by Hubble Space Telescope and International 
Ultraviolet Explorer.}. Moreover, the repeated RM observations of
the same emission line and the RM results of the emission lines with different
ionization in a few AGNs (for instance NGC 5548, 3C 390.3, NGC 3783, 
NGC 7469, Mrk 817 etc.) approximately demonstrated a relation 
$V_{\rm FWHM} \propto \tau^{-1/2}$, showing evidence for potential of SMBHs 
\citep[e.g.,][]{Peterson2000,Peterson2004, Lu2021}, where $V_{\rm FWHM}$ is the 
full-width-half-maximum of emission lines. Considering
that disk-like geometry of BLRs in some AGNs, this relation probably indicates
 nearly Keplerian rotation of the disk BLRs.
More recently, the GRAVITY instrument mounted in Very
Large Telescope Interferometer (VLTI) spatially resolved the BLRs in several AGNs
\citep[e.g., 3C\,273, NGC\,3783, IRAS\,09149 by][respectively]{Sturm2018, 
Gravity2020_09149,Gravity2021_3783} and also
found that their BLRs are approximately characterized by Keplerian rotating disks. 

Moreover, there is growing evidence for 
the existence of sub-structures or the inhomogeneity on the BLR disks
from RM observations. For examples, a well-known phenomenon in RM is that
the emission-line profiles in the mean spectra (corresponding to the entire region of
line emission) and root mean square ones (RMS, to the portion of the
region with response) are generally different in most
AGNs \citep[e.g.,][]{Bentz2009, Denney2009, Barth2013, Du2018MAHA}, which
implies that the BLRs are inhomogeneous in terms of gas distributions.
Furthermore, the sub-features in the velocity-delay maps of NGC~5548
\citep[e.g.,][]{Xiao2018, Horne2021} indicate explicitly that there are probably
gas inhomogeneity in BLRs. In particular, \cite{Horne2021} recently found
helical ``Barber-Pole'' patterns in the Ly$\alpha$ and C {\sc iv} lines of
NGC~5548, which suggests azimuthal structures in their emitting regions.
Multiple-peaked and asymmetric profiles of lines also indicate complicated
BLR structures. These increasing pieces of evidence potentially suggest that
there exist sub-structures in BLRs. Questions naturally arise: are the motions
of the BLR sub-structures chaotic or ordered? What are their physical origin?

Motivated by the observational evidence, 
some preliminary efforts have been made that spiral arms are invoked to
explain the asymmetric double-peaked broad emission lines, however, by only
assuming some analytical form of the arm patterns without considering their
physical origin or introducing any dynamical physics. These efforts were made 
by \cite{Gilbert1999} and \cite{Storchi-Bergman2003, Storchi-Bergman2017} 
who adopted spiral arms to explain the asymmetric double-peaked emission 
lines and their variations in some AGNs. \cite{Horne2004} calculated the 
transfer function of BLRs with arms for RM, but all of them assumed some 
analytical forms of the spirals. All these mathematical models for 
observational data should be derived by the first principle in order to
advance our understanding BLR physics. 

The BLR radii measured by RM span from $10^3 R_{\rm g}$ to $10^5 R_{\rm g}$ 
depending on accretion rates and SMBH masses \citep[see Figure 6 in][]{Du2016V,Du2019},
which overlap with outer part of accretion disks, where 
$\Rg=G\bhm/c^{2}=1.5\times 10^{13}\,M_{8}$\,cm is the gravitational radius, $G$
is the gravitational constant, $c$ is speed of light, and $M_{8}=\bhm/10^{8}\sunm$ 
is the SMBH mass. Some pioneering works suggest that the outer regions  
may be self-gravitating \citep[SG, e.g.,][]{Paczynski1978a,Paczynski1978b} for
viscosity mechanism transferring angular momentum outward, 
in particular, in AGN accretion disks \citep[see details of][]{Shlosman1989}.
It is not difficult to give an rough estimate of the SG
radius using the famous Toomre parameter $Q = \kappa a / {\pi G \sigma}$
\citep{Toomre1964}, which is the criterion of the instability, where $\kappa$ is the
epicyclic frequency (equal to the angular speed $\Omega$ in a Keplerian disk),
$a$ is the sound speed, and $\sigma$ is the
surface density. Using $Q=1$, \cite{Goodman2003} approximated 
$R_{\rm SG}/\Rg=3.1\times 10^{3}\,\alpha_{0.1}^{2/9}\left(L_{\rm Bol}/L_{\rm Edd}\right)^{4/9}
M_{8}^{-2/9}$, where $\alpha_{0.1}=\alpha/0.1$ is the viscosity parameter
and $L_{\rm Bol}/L_{\rm Edd}$ is the Eddington ratio. Interestingly, SG part of the
disks spatially overlap BLRs generally in AGNs, and this motivates that 
BLR structures and dynamics link with SG part somehow.

The ultimate fates of the SG accretion disks remain a matter of debate, however
self-regulation processes, balanced by radiation cooling and the heating
internally from the dissipation of gravito-turbulence
\citep[e.g.,][]{Paczynski1978a, Paczynski1978b, Lin1987, Gammie2001, Lodato2004}
and probably also magneto-rotational instabilities \citep[MRI, e.g.,][]{Balbus1998,
Rafikov2015}, or externally from irradiation (driven by the inner part of accretion
disk, e.g., \citealt{Rice2011, Rafikov2015}) or star formation (inside the
disks, e.g., \citealt{Collin1999, Sirko2003,Wang2011}), are believed to maintain the
disks so that they can stay in marginally stable states. In such states,
non-axisymmetric perturbations (spiral structures) may inevitably grow in the SG
parts, although clumps or stars may also form through condensations if
the cooling time scale $t_{\rm cool} < \beta \Omega^{-1}$, where $\beta$ is a
factor of a few \cite[e.g.,][]{Gammie2001, Rice2003, Rice2011, Kratter2016,
Brucy2021}. The aforementioned
phenomenological evidences of BLR substructures and inhomogeneity enlighten us
that they may be connected with or originated from the spiral arms in the SG
part of accretion disks, at least for those AGNs with disk-like BLRs. 

As the first paper of this series, we adopt the simple
tight-winding approximation and use analytical formulations to discuss the
observational characteristics 
of tightly wound cases of density waves in BLRs. 
The basic formulations, equilibrium states, and boundary
conditions are provided in Section 2. In Section 3, observational features of
the arms are discussed for RM campaigns and interferometric observations of
GRAVITY/VLTI. Brief discussions and conclusions are provided in Section 4 and 5,
respectively. It should be noted that the purpose of this paper is to
demonstrate the general features of BLR spiral arms in observations (e.g., RM)
rather than establish a perfect model.

\section{Model and Formulations}
In the BLR, the ionized gas is rotating with nearly Keplerian velocity around the 
central SMBH. This assumption is supported by the evidence that $V_{\rm FWHM}$ and 
$\sigma_{\ell}$ are roughly proportional to $\sim
\tau_{\rm H\beta}^{-1/2}$ in some AGNs from the
multiple RM campaigns \citep{Peterson2004,Lu2021}, where 
$\sigma_{\ell}$ is velocity dispersion of the H$\beta$ profiles in
the RMS spectra, respectively, and $\tau_{\rm H\beta}$ are
the H$\beta$ time lags with respect to the 5100\AA\, continuum variations. 
However, current accuracy of RM data doesn't allow us to quantitatively 
determine deviations from exactly Keplerian rotation, it should be reasonable 
to assume the disk-BLR has nearly Keplerian. As one of possible mechanisms, 
the magneto-rotational instability (MRI)
\citep[e.g.,][]{Balbus1998} drives radial motion with velocity approximated by 
$u/u_{\rm K} \approx 0.1\alpha_{0.1}$
which is much smaller than the rotation, where $u_{\rm K}$ is the Keplerian roation.
Moreover, for the disks dominated by the central point masses, 
weak self-gravity of the disks can still maintain spiral arms
\citep[e.g.,][]{Lee1999,Tremaine2001}.

In this paper, for simplicity, we apply the classic theory of density waves 
\citep[e.g.,][]{Lin1964, Lin1966,Lin1979} to calculating the broad-emission-line 
profiles and the transfer function for RM, and the differential phase curve 
signal for GRAVITY/VLTI.

\subsection{Self-gravitating disk and BLR}\label{sec:BLR_structure}
Following \cite{Paczynski1978a}, without specifying regulation
mechanisms (dusty gas, star formation, photoionization and accretion etc.), we
use the polytropic relation as a prescription of $Q_{\rm disk}\sim 1$ region for
general cases, which is given by 
%
$p=K_{0}\rho^{1+1/n}$,
%
where $p$ is the pressure, $\rho$ is the density, $K_{0}$ is a
constant, and $n$ is the polytropic index. Fortunately, $K_{0}$ can be generally
constrained by observations of the BLR geometry. The sound speed is given by 
$a_{0}=\left(p/\rho\right)^{1/2}=K_{0}^{1/2}\rho^{1/2n}$. The vertical equilibrium admits
$H=a_{0}\Omega^{-1}$,
and the Toomre parameter is given by 
%
$Q_{\rm disk}={\kappa a_0}/{\pi G\sigma_0}$,
%
where $\sigma_{0}=2\rho H$ is the surface density of the SG region. 
Here we would like to point out that self-gravity is neglected for vertical 
equilibrium for simiplicity. The epicyclic
frequency is given by $\kappa=2\Omega\left(1+\frac{1}{2}d\ln\Omega/d\ln
R\right)^{1/2} \simeq \Omega$, and we adopt
a Keplerian velocity here that $\Omega=\sqrt{G\bhm/R^{3}}\approx 2\times
10^{-9}\,M_{8}^{-1}r_{4}^{-3/2}\,{\rm s^{-1}}$. We have
\begin{equation}
a_0 = \frac{K_0^{1/2}}{(2 \pi G Q_{\rm disk})^{1/2n}}\,\Omega^{1/n},
\end{equation}
\begin{equation}
\sigma_{0} = \frac{2 K_0^{1/2}}{{(2 \pi G Q_{\rm disk})^{(1+2n)/2n}}}\,\Omega^{(1+n)/n},
\end{equation}
and
\begin{equation}
H = \frac{K_0^{1/2}}{(2 \pi G Q_{\rm disk})^{1/2n}}\,\Omega^{(1-n)/n}.
\end{equation}
Given $n$, $K_{0}$ and $Q_{\rm disk}$, we can drive the radial structures of BLR. 

The polytropic index $n$ is a free parameter in this paper, of which the value
can be $0 \sim +\infty$. In practice, for instance, \cite{Paczynski1978a} discussed
the SG disks with  
$n=1.5$ and $3$, \cite{Lubow1998} chose $n=3$ in their work, and
\cite{Korycansky1995} employed $n=5$ as a typical case in the discussion of
axisymmetric waves in accretion disks. In the present paper, the polytropic
index $n$ controls the radius-dependent thickness of the gaseous disk, i.e.,
$H/R \propto R^{(n-3)/2n}$. If $n > 3$, the disk becomes thicker at outer radius
and the surface of the disk tend to be ``bowl-shaped'' (similar to
\citealt{goad2012}). The bowl shape enables the gas on the disk surface to be
illuminated by the ionizing photons from the inner part of the disk, 
otherwise, the geometrically thin disks are not able to be sufficiently ionized.

Observational constrains on BLR gas indicate that
the BLR mass of H$\beta$-emitting gas could be in a wide range from $10^{3}\sim
10^{4}\sunm$, and even more massive \citep{Baldwin2003}. As a simple estimation,
we integrate the surface density of the SG part of accretion disks and obtain a
mass of $M_{\rm disk}\approx 4.1 \times 10^6 \alpha_{0.1}^{-4/5} M_8^{11/5}
\mathdotM^{7/10} r_4^{5/4}\,\sunm$ in light of the standard model of accretion
disks \citep{Shakura1973}, where $\dot{\mathscr{M}}=\dot{M}_{\bullet}/\dot{M}_{\rm Edd}$
is the dimensionless accretion rates, $\dot{M}_{\bullet}$ is the mass accretion rates,
$\dot{M}_{\rm Edd}=L_{\rm Edd}/c^{2}$ is the Eddington rates, and 
$L_{\rm Edd}=1.3\times 10^{46}\,M_{8}\,{\rm erg\,s^{-1}}$ is the Eddington luminosity.
This indicates that the BLRs are the disk surface taking only tiny fraction of
this SG portion. In the present paper, we assume that the ionized BLR gas is proportional 
to the density of the disk 
$\rho_{\rm ion} \propto \rho$ (see the following Section \ref{sec:line_profiles}).
It is found that $M_{\rm disk}/\bhm\ll 1$, namely, the accretion disks
are much lighter than the central SMBH unless for cases with extremely super-Eddington 
accretion rates ($\mathdotM\gg 1$). Though $\mathdotM\sim 900$ AGNs have been found 
from RM campaigns \citep{Du2014,Du2016,Du2016V,Du2018}, we limit the present scope 
for sub-Eddington accretion disks ($\mathdotM\lesssim 3$) with nearly Keplerian 
rotation of a point potential of SMBH mass.

We would point out here a possibility that enough high density outflows 
emitting high ionization lines could partially shield the outer part of BLRs 
making reverberation of H$\beta$ line complicated \citep[][]{Dehghanian2021}. 
If it happens in such a case, H$\beta$ line would undergo
a holiday driven by the inner outflows (of \civ\ line) like in NGC 5548 
\citep[see Figure 7 in][]{Pei2017}, namely, appearing a lack of reverberation.
However, H$\beta$ line only had the holiday once over the last 20 
years \citep{Pei2017}, indicating that such a 
holiday is quite rare and the obscurations are not common.

\subsection{Equations and boundary conditions}
\subsubsection{Perturbation equations}
We adopt the formalisms and notations used in \cite{Lin1979}. For
an $m$-fold axisymmetric perturbation,
$(u,\vup,\sigma)=(u_{0},\vup_{0},\sigma_{0})+(u_{1},\vup_{1},\sigma_{1})e^{i(\omega
t-m\varphi)}$, where $u$ and $v$ are the radial and azimuthal velocities, and
$\sigma$ is the surface density. The parameters with subscript ``0'' correspond
to the equilibrium state (we take $u_{0}=0$ in this paper), which are functions
of radius. We have the perturbation of the equations given in Appendix
\ref{app:basic_equations},
\begin{equation}
\frac{1}{R}\frac{d}{dR}(R\sigma_{0}u_{1})-\frac{im}{R}\sigma_{0}\vup_{1}+i(\omega-m\Omega)\sigma_{1}=0,
\end{equation}
\begin{equation}
i(\omega-m\Omega)u_{1}-2\Omega \vup_{1}=-\frac{d(\psi_{1}+h_{1})}{dR},
\end{equation}
and
\begin{equation}
\frac{\kappa^{2}}{2\Omega}u_{1}+i(\omega-m\Omega)\vup_{1}=im\frac{\psi_{1}+h_{1}}{R},
\end{equation}
yielding the following differential equation
\begin{equation}\label{eq:h1}
\frac{d^{2}}{dR^{2}}(h_{1}+\psi_{1})+\calA\frac{d}{dR}(h_{1}+\psi_{1})+
\calB(h_{1}+\psi_{1})=-\calC h_{1},
\end{equation} 
where $h_{1}=a_0^{2}\sigma_{1}/\sigma_{0}$. Here the
coefficients $\calA,\calB,\calC$ are given in Appendix \ref{app:coefficients}.

The Poisson equation of the SG portion of the accretion disks reads 
$\nabla^{2}\psi=4\pi G \sigma \delta(z)$, through vertical integration, yielding
\begin{equation}\label{eq:Poisson}
\frac{d\psi_{1}}{dR}=-\frac{\psi_{1}}{2R}-is_{\rm k}\Sigma h_{1},
\end{equation}
where $\delta(z)$ is the Dirac-$\delta$ function,
$\Sigma=2\pi G\sigma_{0}/a_{0}^{2}$, and $s_{\rm k}=\mp 1$ is the sign function of wave vector 
$(\vec{k})$ for trailing and leading waves, respectively.  
The Poisson equation holds approximately
in the order of $(H/R)^{2}$. Combining the perturbation equations, we have the equation of the 
reduced enthalpy ($U$)
\begin{equation}\label{eq:U}
\frac{d^{2}U}{dR^{2}}+k_{3}^{2}U=0,
\end{equation}
where 
\begin{equation}
U=h_{1}\left[\frac{\kappa^{2}(1-\nu^{2})}{\sigma_{0}R}\right]^{-1/2}
       \exp\left(\frac{i}{2}\int\Sigma dR\right),  
\end{equation}
and
\begin{equation}
k_{3}^{2}=\left(\frac{\kappa}{a_{0}}\right)^{2}\left(Q_{\rm disk}^{-2}-1+\nu^{2}\right);\quad 
\nu=\frac{\omega-m\Omega}{\kappa}.
\end{equation}
\cite{Bertin2014} presents more detailed derivations of the above equations (\ref{eq:h1} and \ref{eq:U}). 
Equation (\ref{eq:U}) works approximately in the order of $H/R$, which agrees with that of the Poisson 
equation. As a first application of density waves in BLR, we retain this order of approximation for 
simplicity.

\subsubsection{Boundary conditions}
In the context of spiral galaxies, the outer boundary conditions are imposed by
radiation condition \citep{Lin1979}. Considering that dynamics of dusty 
and dust-free gas will be very different due to radiation pressures (either from
local or central part) of accretion disks. The boundary could be distinguished by
the dust sublimation radius. For the present SG disk, the outer boundary
is fixed at the inner edge (inward is dust-free) of dusty torus where 
the waves are evanescent.
Although the dusty torus is generally not spatially resolved (except for NGC
1068 which shows a near-infrared cavity with a sharp edge in
\citealt{Gravity2020}), fortunately, the RM campaigns of near-infrared continuum
emissions show that the inner edge of torus is about $R_{\rm torus}\approx 0.1\,
L_{43.7}^{0.5}\,{\rm pc}$ \citep{Suganuma2006, Koshida2014, Minezaki2019,Lyu2019},
where $L_{43.7}$ is the $V$-band luminosity in units of 
$10^{43.7}\ergs$ \citep[see the latest version in][]{Minezaki2019}. 
It was long believed that the out edge of BLR is just the 
inner edge of torus \citep[e.g.,][]{Netzer1993, Suganuma2006, Czerny2011}. 
Therefore, we can obtain the outer radius of the BLR
\begin{equation}\label{eq:Rtorus}
R_{\rm out}=205.9\,\eta_{0.1}^{1/2}\epsilon_{10}^{-1/2}\mathdotM^{1/2}M_{8}^{1/2}\,{\rm ltd},
\end{equation}
from $R_{\rm out}=R_{\rm torus}$, the bolometric luminosity is
$L_{\rm Bol}=\epsilon L_{V}=\eta \dot{\mathscr{M}}\dot{M}_{\rm Edd}c^{2}$,
$\epsilon=10\epsilon_{10}$ is the bolometric luminosity correction factor and $\eta=0.1\eta_{0.1}$
is the radiative efficiency. This agrees with the 
observation of NGC 1068 \citep{Gravity2020} by GRAVITY onboard 
VLTI. We adopt a vanishing perturbation of density at 
the outer boundary.
For the simplest case of the inner boundary, we assume
\begin{equation}\label{eq:sigma1}
\frac{dU}{dR}=0,
\end{equation}
just as in \cite{Lau1978}. We set the inner radius to be 10 percent of the outer
radius, and have checked that the detailed value of inner radius does not change
the general features of spiral arms. With the boundary conditions,
it becomes a eigenvalue problem to solve Eqn (\ref{eq:U}). Then, we can obtain the
perturbation of the surface density \citep[more details can be found in,
e.g.,][]{Lau1978, Lin1979}.

It should be pointed out that outer boundary conditions could be revised for
individual AGNs if the spatially resolved conditions are different from the 
present. The adopted conditions are obvious that any H$\beta$ photons will be 
extincted by dusty gas within the torus whatever it is a kind of outflows
\citep{Konigl1994,Elitzur2006}, clumpy structures \citep{Nenkova2008} or 
classical continue torus \citep{Antonucci1993}. On the other hand, accretion 
disks could extend outward and correspond to the
mid-plane of torus, and density waves also extend (but depends on the local
radiation pressure).  In some AGNs, ALMA
(Atacama Large Millimeter/submillimeter Array) observations show spiral arms, 
like in a few Seyfert galaxies \citep{Combes2019}, and it would be
interesting to test if they are consistent with ones in BLRs. 
Recent interferometric observations of NGC 1068 show much more complicated 
structure \citep{Rosas2022} as well as the counter-rotating disk from 0.2 
to 7 pc by ALMA observations \citep{Imanishi2018,Impellizzeri2019}.
In this paper, the simplest conditions are taken for the outer boundary physics.

\subsection{Line profiles and 2-dimensional transfer functions}
\label{sec:line_profiles}
The emissivity distributions in BLRs are still unclear from observations. From
photoionization, the locally optimally emitting clouds model suggests that the
line emission irradiates the most efficiently from a
relatively narrow range of ionization parameter
$U_{\rm ion} = Q_{\rm ion} / 4 \pi R^2 c n_{\rm ion}$, where $Q_{\rm ion}$ is the
number of ionization photons, $n_{\rm ion} = \rho_{\rm ion} / m_{\rm H}$ is the
number density of hydrogen, $\rho_{\rm ion} \propto \rho =
\left[\sigma_{0}(R)+\sigma_{1}(R,\varphi)\right] / 2H =
\left[\sigma_{0}(R)+\sigma_{1}(R,\varphi)\right] \Omega / 2 a_0$ is the ionized
hydrogen density and assumed to be proportional to the density of the disk, and
$m_{\rm H}$ is the mass of hydrogen atom \citep{Baldwin1995, Korista1997,
Korista2000}. For simplicity, we assume that the emissivity (reprocessing
coefficient) is a Gaussian function of ionization parameter as
\begin{equation}\label{eq:Xi}
  \Xi_{R} \propto \frac{1}{\sqrt{2\pi} \sigma_U} e^{-(U_{\rm ion} - U_{\rm c})^2 / 2 \sigma_U^2},
\end{equation}
where $U_{\rm c}=U_{\rm ion}(R_{\rm c})$ is the ionization parameter
corresponding to the most efficient line emission at radius $R_{\rm c}$ and
$\sigma_U = \tilde{\sigma}_U \times (U_{\rm ion,max} - U_{\rm ion,min})$ is a
parameter that controls the range of line emission. Actually, the form of Eqn. 
(\ref{eq:Xi}) is one simplified version of the popular model 
\citep[e.g.,][]{Pancoast2014,Li2018}. The typical BLR radii are on
average smaller than the inner edges of tori by factors of $4\sim5$
\citep{Koshida2014, Minezaki2019}, thus we adopt $R_{\rm c} = 1/4 R_{\rm torus}$
and assume $\tilde{\sigma}_U = 0.20$ (corresponding to a not very compact
line-emitting region).

Given the configuration of the disk-like BLR in
Section \ref{sec:BLR_structure}, the emission-line profile can be expressed as
\begin{equation}
F_{\ell}(\lambda)=\int_{R_{\rm in}}^{R_{\rm out}}RdR\int_{0}^{2\pi}\,d\varphi\, \Xi_{R}\,
            \delta\left[\lambda-\lambda_{0}\left(1+\frac{\bm{\vup}\cdot\bmnobs}{c}\right)\right],
\end{equation}
where $\bm{\vup}$ is the velocity of emitting gas, $\bmnobs=(0,\sin i_{0},\cos i_{0})$ 
is the vector of line of sight, and $i_{0}$ is the inclination angle.
\cite{Blandford1982} developed the linear reverberation technique to map BLRs. 
Denoting the ionizing continuum light curve and broad-line light curve at velocity $v$ of the 
line profile as $L_{\rm c}(t)$ and $L_{\ell}(\vup,t)$, respectively, we have
\begin{equation}
L_{\ell}(\vup,t)=\int_{-\infty}^{\infty}dt^{\prime}\,L_{\rm c}(t^{\prime})\Psi(\vup,t-t^{\prime}),
\end{equation}
where $\Psi(\vup,t)$ is the 2D transfer function (velocity-delay map), $\Psi(\vup,t)=0$ for $t<0$ and 
$\Psi(\vup,t)\ge 0$ for $t\ge 0$. RM campaigns obtain $L_{\ell}(\vup,t)$ and $L_{\rm c}(t)$, which can 
be used to infer $\Psi(\vup,t)$. With the geometric 
and kinematic configurations of BLRs with density waves, the 2D transfer function can be obtained by
\begin{equation}
\Psi(\vup,t)=\int d{\bmR}\,\frac{g({\bmR},\vup)\,\Xi_{R}}{4\pi R^{2}}\,
          \delta\left[t-\frac{(R+{\bmR}\cdot{\bmnobs})}{c}\right], 
\label{eqn_psi}
\end{equation}
where $g({\bmR},\vup)$ is the projected one-dimensional velocity distribution function. 
With Equation~(\ref{eqn_psi}), features of density waves can be calculated and then compared 
with observations.

In principle, high fidelity data from RM campaigns can be used to generate 2D transfer
functions through the maximum entropy \citep{Horne2004} or the improved Pixon-based method \citep{Li2021}.
Spiral arms as prominent inhomogeneous components of BLRs can be directly tested avoiding 
uncertainties of explanations in light of complicated profiles alone.

\begin{figure*}
    \centering
    \includegraphics[width=0.95\textwidth]{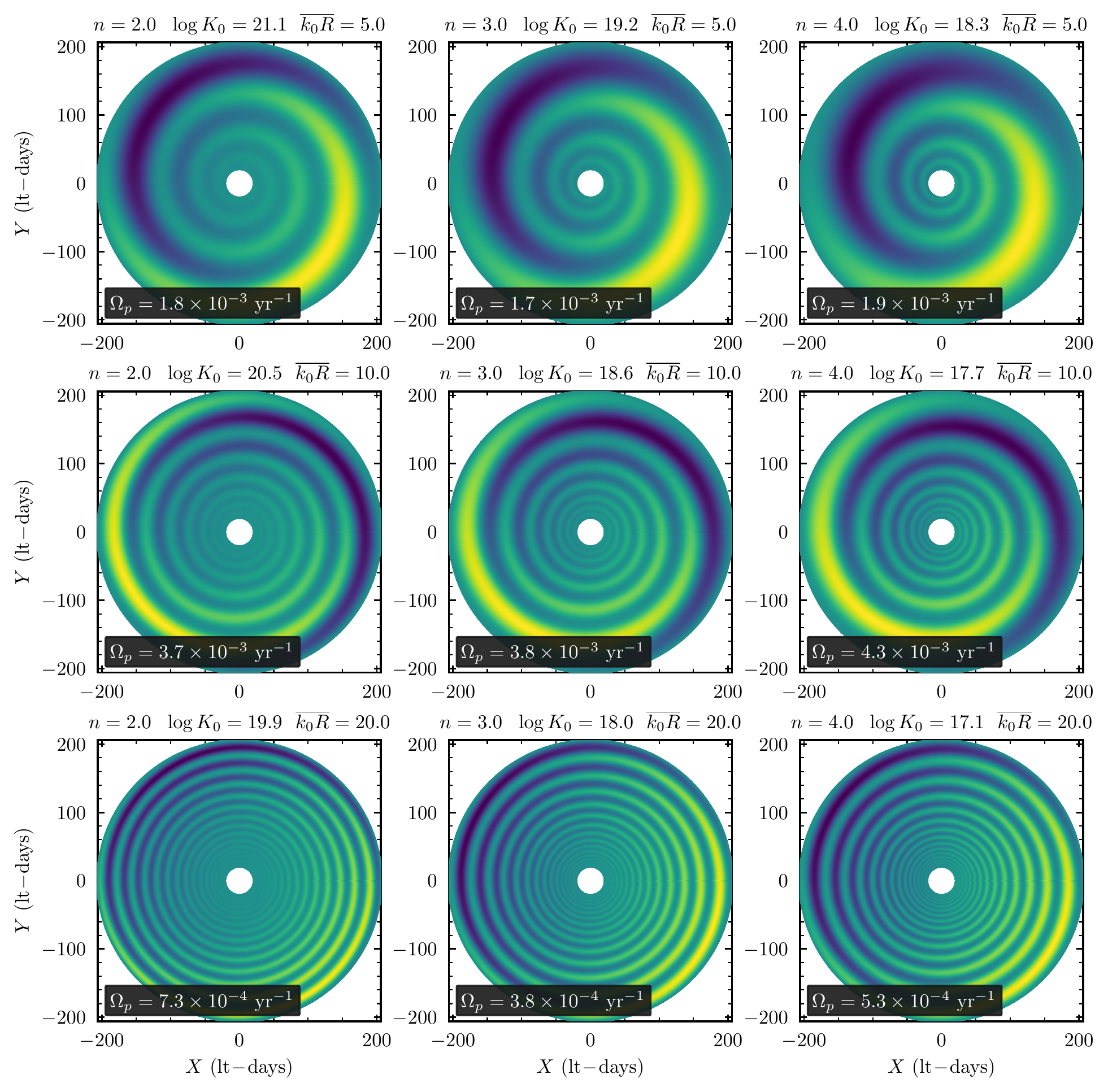}
    \caption{The $m=1$ mode perturbations $\sigma_{1}/\sigma_{0}$
    for different parameters of equilibrium states. It generates one arm
    structure appearing as a crest (with colors from green to yellow) on an
    axisymmetric disk. The bright arm is undergoing reverberation with response
    to variations of ionizing continuum. The dark arm shows a density valley
    without or weakly response to the continuum variations. $\Omega_{p} = \omega/m$ is 
    the rotation speed of the arm pattern.}
    \label{fig:spiral-arm}
\end{figure*}

\subsection{Signals for GRAVITY/VLTI}
As a powerful technique, spectroastrometry (SA) developed from ``Differential Speckle Interferometry" 
\citep{Becker1982,Petrov1989,Rakshit2015} measures the center of photons and therefore greatly improve 
the spatial resolution. The spectroastrometry with GRAVITY/VLTI can reach 
an unprecedentedly high spatial resolution of $\sim 10\mu$as. It had been successfully applied 
to 3C 273 for the geometry and kinematics of its BLR \citep{Sturm2018}\footnote{Current accuracy of
GRAVITY measurements of 3C 273 is only enough to test a simple disk model rather than sub-structures
of spiral arms \citep[see details of ][]{Sturm2018}.}. 
The spiral arms developed 
from density waves may show some signatures that can be detected by
GRAVITY/VLTI. The detailed scheme for spectroastrometry technique is described 
in \cite{Rakshit2015} and \cite{Songsheng2019}. Below we outline a brief description for the sake 
of completeness. Given the surface brightness distribution, we have
the photon center of the source at wavelength $\lambda$
\begin{equation}\label{eq:ph-center}
\bm{\epsilon}(\lambda) = 
\frac{\int \bm{\alpha} \calO(\bm{\alpha},\lambda)  \dd[2]{\bm{\alpha}}}
{\int \calO(\bm{\alpha},\lambda) \dd[2]{\bm{\alpha}}},
\end{equation}
where $\calO(\bm{\alpha},\lambda)=\calO_{\ell}+\calO_{\rm c}$ is the surface brightness 
distribution of the source contributed by the BLR and continuum emissions, and
$\bm{\alpha}$ is the angular displacement on the celestial sphere. Given the geometry and 
kinematics of a BLR, its $\calO_{\ell}$ can be calculated for the broad emission line 
with the observed central wavelength $\lambda_{\rm cen}$ through
\begin{equation}\label{eq:calO}
\calO_{\ell}=\int \frac{\Xi_R F_{\rm c}}{4 \pi R^2} f(\bm{R},\bm{\vup})\,
             \delta\left(\bm{\alpha}-\bm{\alpha}' \right)\delta\left(\lambda -\lambda^{\prime}\right)
             \dd[3]{\bm{R}} \dd[3]{\bm{\vup}},
\end{equation}
where $\lambda^{\prime}=\lambda_{\rm cen}\gamma_{0}\left(1+\bm{\vup}\vdot\nobs/c\right)
\left(1-R_{\rm S}/R\right)^{-1/2}$ includes the gravitational shift due to the central 
SMBH, $R_{\rm S}=2R_{\rm g}$ is the Schwarzschild radius,
$\gamma_{0}=\left(1-\vup^2/c^2\right)^{-1/2}$ is the Lorentz factor, 
${\bm \alpha}^{\prime}=\left[\bm{R}-\left(\bm{R}\vdot\bmn\right)\bmn\right]/D_{\rm A}$,
$\bm{R}$ is the distance to the central SMBH, $f(\bmR,\bmv)$ is the velocity distribution 
of BLR clouds at $\bmR$, and $F_{\rm c}$ is the ionizing flux.
By introducing the fraction of the emission-line flux to the total ($\fline$), we have
\begin{equation}\label{eq:epsilon}
\bm{\epsilon}(\lambda) = \fline\,\bm{\epsilon}_{\ell}(\lambda),
\end{equation}
where
\begin{equation}
\bm{\epsilon}_{\ell}(\lambda) = \frac{\int \bm{\alpha} \calO_{\ell}  
   \dd[2]{\bm{\alpha}}}{\int \calO_{\ell} \dd[2]{\bm{\alpha}}},\,\,\,\fline 
= \frac{\Fline(\lambda)}{F_{\rm tot}(\lambda)},\, \ \ \Fline(\lambda) 
= \int \calO_{\ell} \dd[2]{\bm{\alpha}},
\end{equation}
and
\begin{equation} 
   F_{\rm tot}(\lambda)=\Fline(\lambda) + F_{\rm c}(\lambda). 
\end{equation}
For an interferometer 
with a baseline $\bm{B}$, a non-resolved source, with a global angular size smaller than 
its resolution limit $\lambda / B$, has the interferometric phase
\begin{equation}\label{eq:phase}
\phi_*(\lambda,\lambda_{\rm r})=-2\pi\bm{u}\vdot[\bm{\epsilon}(\lambda)-\bm{\epsilon}(\lambda_{\rm r})],
\end{equation}
where $\bm{u}=\bm{B}/\lambda$ is the spatial frequency, and $\lambda_{\rm r}$ 
is the wavelength of a reference channel. Given the geometry and kinematics of the BLR,
the spectroastronometric signals can therefore be calculated.

\begin{figure*}
    \centering
    \includegraphics[width=0.95\textwidth]{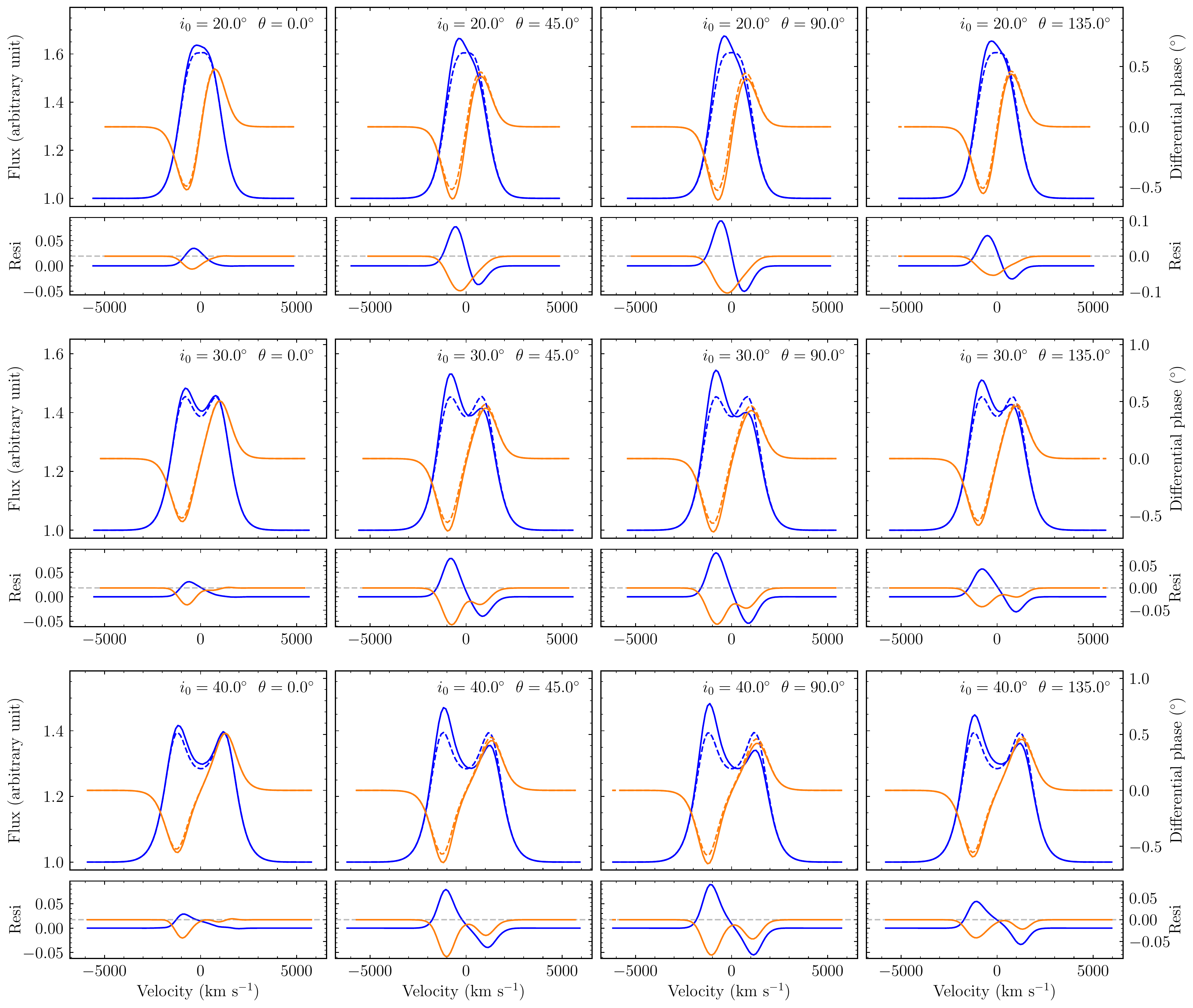}
    \caption{Characterized profiles of BLR-disk ($\sigma_{0}+\sigma_{1}$) with
    one-arm structure (in blue) and differential phase curves (in orange). The
    panels are for different orientation and azimuthal angles. Dashed-lines
    represent emission from an homogeneous disk, and the solid lines from the
    disk with spiral-arm. In the calculations, we take the perturbations of
    surface density $(\sigma_1/\sigma_0)_{\rm max} = 0.2$. The
    differences between the solid and dashed lines are provided as residuals
    with the corresponding colors. The differential phases can be measured by
    GRAVITY, and the residual to the homogenous disk can be also detected for
    the BLR with the present parameters. Less massive black holes can be
    measured by GRAVITY+ (the next generation of GRAVITY).}
\label{fig:profile}
\end{figure*}

\begin{figure*}
    \centering
    \includegraphics[width=0.95\textwidth]{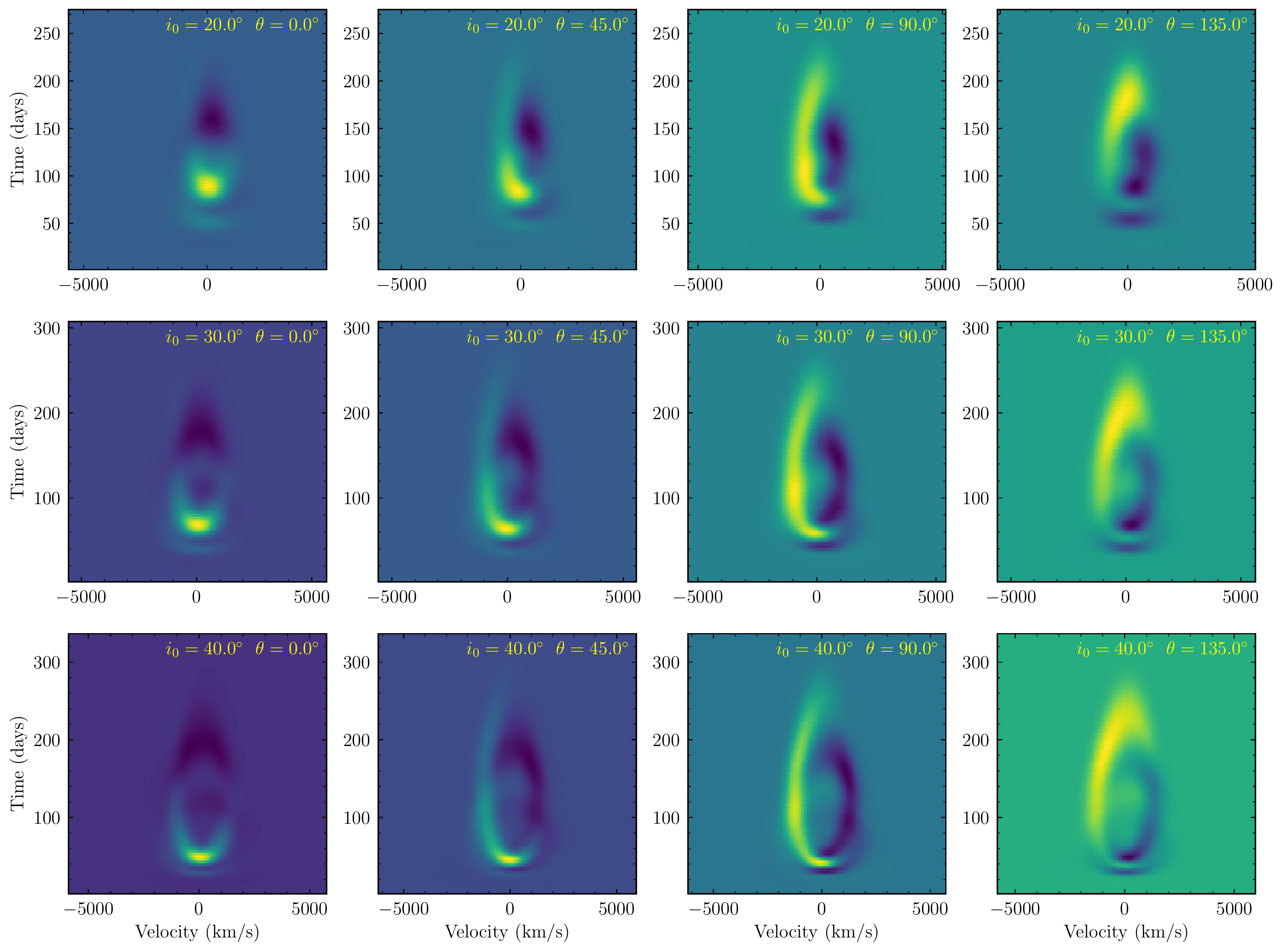}
    \caption{
    Velocity-delay maps [$\Psi(v,t)$] of the spiral arms for 
    the case with $n=4$ and $\overline{k_0 R}=5$.
    For clarity, we only plot
    the response of the density perturbations ($\sigma_{1}$). The color codes
    the strength of $\Psi(v,t)$ (brighter for stronger and darker for weaker
    responses). }
\label{fig:BLR}
\end{figure*}

\section{Results: SA and RM signals}
\label{sec:results}
It is known that the nearly-Keplerian disks dominated by the potential of central 
sources favor $m=1$ mode, namely a single spiral arm \citep{Adams1989,Shu1990,Lee2019}.
We can obtain  $\omega$ from the
eigenvalue problem of Equation (\ref{eq:U}) as well as the perturbed component 
($\sigma_{1}/\sigma_{0}$) for density waves. In this paper, we 
only focus on the general patterns of the arms for the nearly Keplerian 
rotating disks, and we leave their growth 
rates connected with the imaginary part of $\nu=\left(\omega-\Omega\right)/\kappa$
in the second paper of this series (Du et al. 2022 in preparation).

The dispersion relation can be used to give a rough estimate to the tightness of 
winding of the spiral arms, which is expressed by 
$(\omega-\Omega)^{2}=\kappa^{2}+k^{2}a_{0}^{2}-2\pi G|k|\sigma_{0}$  \citep{Lin1979}, 
we have
\begin{equation}
k=-k_{0}\left[1\pm\sqrt{1-Q_{\rm disk}^{2}(1-\nu^{2})}\right],\quad
k_{0}
     =\frac{(2 \pi G)^{1/2n} Q_{\rm disk}^{(1-2n)/2n}}{K_0^{1/2} \Omega^{(1-n)/n}}.
\end{equation}
Pitch angles of spiral arms are determined by $\tan i=1/kR$ for given $n$, $K_{0}$ 
and $Q_{\rm disk}$, and $k_{0}R$ can be representative of the global pitch angles. 
In order to conveniently show the spiral arms, we use the proxy of pitch angles 
defined by
%
$\overline{k_0 R}=\left(R_{\rm out}-R_{\rm in}\right)^{-1}
         \int_{R_{\rm in}}^{R_{\rm out}} k_{0}R\, dR$
%
along the radial axis as an input parameter in following calculations, rather than 
$K_0$. Moreover, the validity of the tightly wound approximation can be obviously 
justified by this parameter. Given $\mathdotM$ and $\bhm$, we show dependence of 
$\overline{k_0 R}$ on $K_0$ and $n$ in Appendix \ref{app:wavenumbers}. In general,
decreasing $K_0$ and $n$ leads to increases of $\overline{k_0 R}$, and 
$\overline{k_0 R}$ slightly depends on $\mathdotM$ 
and $\bhm$. This is caused by the dependence of inner 
and outer boundaries on $\mathdotM$ and $\bhm$, see Eqn \ref{eq:Rtorus} and Appendix 
\ref{app:wavenumbers}. As a representative, we focus on the case with 
$\mathdotM = 1.0$ and $\bhm = 10^8 M_{\odot}$, and $Q_{\rm disk}=1$ in 
the present paper.

Figure \ref{fig:spiral-arm} shows the spiral arms of several cases with different 
$K_0$ and polytropic index $n$ for $m=1$ mode. In this Figure, we arrange the panels 
in the same line have the same $\overline{k_0 R}$. The corresponding
eigenvalues ($\Omega_{p}=\omega/m$) are marked in the left corners in individual
panels. Although the BLR gas may not be illuminated by the central ionizing
radiation if $n < 3$ (see more details in Section \ref{sec:BLR_structure}), we
still show the cases with $n=2$ in Figure \ref{fig:spiral-arm} for comparison.
It is obvious that the arms wind more tightly if $\overline{k_0 R}$ is larger.
In addition, the contrast of the perturbations in the central regions becomes clearer 
if $n$ increases. 

Figure \ref{fig:profile} shows the emission-line profiles of the
case with $n=4$ and $\overline{k_0 R}=5.0$ (the arm pattern in the upper-right 
corner {\ccyan in Figure~\ref{fig:spiral-arm}}) for different
azimuthal and inclination angles ($\theta$ and $i_{0}$) of the line of sight.
The maximum value of $\sigma_1/\sigma_0$ is fixed to be
$(\sigma_1/\sigma_0)_{\rm max} = 0.2$. The real situations may be smaller or
larger than the value adopted here. The azimuthal angle $\theta=0^{\circ}$
refers to the line of sight that the observer looks the BLR 
pattern in Figure \ref{fig:spiral-arm} from the right side, and the azimuthal
angle increases counter-clockwise (the rotational velocity of gas
is also counter-clockwise). For comparison, the profiles of the corresponding
unperturbed disks without spiral arms are also superposed in
Figure~\ref{fig:profile}. 
As expected, the unperturbed disk (geometrically thin) structure of the BLR
generates symmetric double-peaked profiles. The double peaks are
blended if the disk tends to be observed from a face-on direction (e.g.,
$i_0=20^{\circ}$ in Figure \ref{fig:profile}). The line profiles of the disks
with the spiral arms are obviously asymmetric. For example, in the cases with
$\theta=45^{\circ}$, the blue peaks are significantly higher than the red ones
because the bright parts of the arms are approaching. 
As $\theta$ increases from $0^{\circ}$, the asymmetry first becomes stronger
until $\theta\sim45^{\circ}-90^{\circ}$ and then gets weaker. The
width of the line profile increases if the inclination angle increases.
The differences between the profiles with and without spiral arms
are also shown in Figure \ref{fig:profile}. {\ccyan If $\sigma_1/\sigma_0$ is larger, 
the asymmetry of the profiles will become stronger, and vise versa.}

In Figure \ref{fig:profile}, we also calculate spectroastrometric signals
(orange lines), which are detectable for GRAVITY/GRAVITY+ on
VLTI. The signals are significantly different from the standard $S$-curves. The
$S$-curves of the unperturbed BLR disks are symmetric that the amplitudes of the
blue and red peaks are the same. The phase curves of the BLR with the spiral arm
for different azimuthal and inclination angles are also asymmetric just like
their emission-line profiles. The differences between the perturbed and
unperturbed BLR disks are also shown as the residuals in Figure
\ref{fig:profile}. They are the functions of inclination and azimuthal angles.
Comparing with the unperturbed BLR disk, the amplitudes of the blue troughs in the
cases with the spiral arm are stronger and those of the red peaks become weaker.
Similar to the line profiles, the asymmetry of the phase curve first increases
if $\theta$ becomes larger, and then decreases after 
$\theta\sim45^{\circ}-90^{\circ}$. The width of the phase curve increases with 
the inclination angle
increasing. The current GRAVITY/VLTI can conveniently detect differential phase
angles to $\sim 0.1^{\circ}$, and GRAVITY+ as its next generation\footnote{see
more details from \url{https://www.mpe.mpg.de/7480772/GRAVITYplus_WhitePaper.pdf}}
will definitely observe the features much weaker than the present in the near
future.

Figure \ref{fig:BLR} shows the transfer functions (velocity-delay maps) of the
spiral arm with $n=4$ and $\overline{k_0 R}=5$ for different
azimuthal and inclination angles. 
For a brief comparison with the present, we refer readers to the figures showing
the transfer functions of Keplerian disks in \cite{Welsh1991} and \cite{Wang2018}. 
A Keplerian disk shows a symmetric bell-like feature. For clarity,
in Figure \ref{fig:BLR}, we only plot the response of the density perturbations
($\sigma_{1}$). For comparison, an example of the transfer function for
the unperturbed disk is shown by Figure \ref{fig:vdm_unperturbed} in Appendix
\ref{app:transfer_function_unperturbed}. As shown in Figure \ref{fig:BLR}, the
major influences of the spiral arm are the variations of the bell's waist. We
find that the higher and lower density perturbations (wave crest and wave
trough, see also in Figure \ref{fig:spiral-arm}) generate
positive and negative signals (stronger and weaker responses) with superposition
to the transfer functions of Keplerian disks, respectively. Actually, there is 
some deficit of the right waist in the bell-like transfer function of NGC 5548 as 
shown in \citep{Xiao2018}. When $\theta=0^{\circ}$, the positive signals are mainly
located in near side (smaller time lags) and the negative signals
tend to be at larger time lags. Along with  $\theta$ increasing
from $0^{\circ}$, the positive and negative signals rotate clockwise. Measuring
the rotation of the features is one of the keys to test the presence and
dynamics of the spiral arms in BLRs.

For cases with $Q_{\rm disk}=1+\Delta Q$, where $\Delta Q<0.3$, 
we have done the calculations and found the general patterns of arms
do not change significantly for given $\overline{k_{0}R}$. We omit the figures.   

\section{Discussions}
\subsection{Tight-winding approximation}\label{sec:complex_BLR}
Employing the traditional WKB approximation \cite[e.g.,][]{Lin1979}, 
we for the first time apply the theory of density waves to BLRs for
broad emission line profiles, differential phase curves, and
velocity-delay maps. The validity of this approximation can be simply estimated by
comparing the global pitch angle proxy of $\overline{k_{0}R}$  
as shown in Figure \ref{fig:spiral-arm} with detailed calculations
in Du et al. (2022 in preparation) which relaxes the WKB approximation for the more loosely wound cases. We find that the difference of pitch angles 
can be less
than $\lesssim 20\%$ for $\overline{k_{0}R}=5$. It is generally
believed the WKB approximation is good enough for $\overline{k_{0}R}\gtrsim 5$, which
is consistent with \cite{Lin1979}. 

Moreover, the non-linear effects have been extensively studied by \cite{Lee1999},
who draw a conclusion that single-armed density waves can exist even in nearly Keplerian disks with only
weak self-gravity. Pitch angle of the arms can be significantly 
larger with non-linear effects. Relaxing the WKB approximation schemed by \cite{Adams1989}
will generate more general results for AGN BLR issues as shown in a 
forthcoming paper (Du et al. 2022 in preparation). On the other hand, some simulations show that
$m=1$ is also favored when the disk mass is comparable with the central SMBHs
\cite[e.g.,][]{Lodato2004, Kratter2016}. In such a context, the disk will be thicker than the 
present cases so that the Poisson equation has more complicated expression than 
Eqn. (\ref{eq:Poisson}). Density waves will be modified by radial self-gravity of
the disks.

\subsection{Observational appearance}
The present profiles of broad emission lines can be conveniently compared 
with observations.  The suggested spiral arms for asymmetric profiles of some AGNs 
\citep[e.g.,][]{Eracleous1994,Storchi-Bergman2017} could originate from the
results of self-gravity instability. Observational appearance should be tested by
examining individual AGNs and statistic properties.
Asymmetric profiles of
Palomar-Green quasars are common, and statistically, the asymmetries are significantly
correlated with the strengths of \Feii\, (which clearly depends on accretion
rates $\mathdotM$; see Figure 5 in \citealt{Boroson1992}). This has a clear
implication that the homogeneity of ionized gas distributions is governed by
accretion rates. \cite{Marziani2003, Marziani2009, Marziani2010} investigated
the line profiles of low- and intermediate-redshift AGNs, and concluded
that the line asymmetry changes systematically along the so-called quasar 
``eigenvector 1 sequence''. Parameters of $n$, $K_{0}$, and $Q_{\rm disk}$ 
may dependent directly on $\mathdotM$ in reality, resulting in the 
dependence of the arms on accretion rate. This could probably explain these 
phenomena. We also note that asymmetries of H$\beta$ profiles are changing
with time from red to blue asymmetries or {\ccyan reverse}, which could be
naturally explained by the pattern motion of $m=1$ mode spiral arms. 
High fidelity RM of AGNs, that employs high spectral resolution and
homogeneous cadence, will finally reveal sub-structures of the
BLRs through detailed 2D transfer functions from response to the ionizing
sources \citep{Welsh1991,Horne2004,Wang2018,Songsheng2019,Songsheng2020} as 
well as the spectroastrometric observations of GRAVITY/VLTI with the predicted 
characteristics. Detailed comparisons with observations will be deferred to 
a future paper.

However, the true situation can be more complicated than the simplified
model adopted here, which could make the density waves (spiral arms) more
complex, especially in the cases of radiation pressure-driven warped disks
\citep{Pringle1996}, MRI-driven turbulence or other instabilities \citep[e.g., a
brief review in][]{Ogilvie2013}, or star formation \citep{Shlosman1989,
Collin1999, Gammie2001, Collin2008, Wang2011, Wang2012}. Magnetic fields
could be very important in some cases. Moreover, fast cooling
could make the disk suffer from violent instability, condense into discrete clouds, 
and even generate filamentary spiral pattern
\cite[e.g.,][]{Gammie2001, Rice2003, Rice2011, Kratter2016, Brucy2021}. However,
on one hand, heating caused by MRI or radiation from inner region may balance
the cooling effect. On the other hand, from the perspective of observation,
high-resolution spectroscopy in \cite{Arav1997, Arav1998} provides a lower limit
to the number of clouds in BLRs and exclude that BLRs consist of ``discrete''
clouds. Whatever how complicated the physics is in BLRs, future detections will 
advance understanding the mystery of the BLRs.

\subsection{Relation between BLR and accretion disk}\label{sec:BLR-disk}
The origin of BLRs and their relation with accretion disks are still a puzzle.
In the present paper, we assume that the BLR is the illuminated surface layer of
the accretion disk (in the SG region). Such kind of assumption was also adopted
by, e.g., \cite{goad2012}. Only in $n>3$ cases, the gas on the
surface can be illuminated by the central ionizing photons. It should be noted
that radiation pressure from the disk may puff up the height of this region
\citep[e.g.,][]{Emmering1992, Murray1995, Chiang1996, Czerny2011, Elvis2017,
Baskin2018} and provide a covering factor that is large enough to explain the
BLR observations, which will ease the restriction of polytropic index. However,
more thick disk may potentially stabilize against non-axisymmetric perturbation
and reduce the lifetime of spiral arms \citep{Ghosh2021}. The observation in
\cite{Horne2021} has indicated the azimuthal structures in BLRs, which gives a
constraint that the lifetime of the arms cannot be too short. The influence of
the thick BLR layers should be investigated both theoretically and
observationally in future. 

\subsection{Self-gravity of the accretion disks}
To maintain density waves needs $Q_{\rm disk}\sim 1-1.3$ \citep[e.g.,][]{Lodato2004} 
driven by several mechanisms mentioned in Section 1, however, it is expected to distinguish
them from observations. Star formation in the
self-gravitating disks could support the state of $Q_{\rm disk}$ as suggested by 
\citep[e.g.,][]{Shlosman1989,Collin1999,Thompson2005,Wang2011}. As a self-regulation, 
higher star formation will decrease the surface density of the disks increasing 
$Q_{\rm disk}$ whereas
lower star formation rates increase the surface density decreasing $Q_{\rm disk}$.
Except for releases of gravitational energy of accreting gas, star formation and supernovae
explosion will supply additional energies to this region. As independent evidence for this,
fortunately, AGNs and quasars are known to be metal-rich providing observational constraints 
on these processes. It would be interesting to test potential dependence of broad emission
line profiles on the metallicity ($Z$). As hint evidence, asymmetries of H$\beta$ profiles 
strongly correlate with \Feii\, strength (${\cal R}_{\rm Fe}$) 
\citep[see Figure 5 in][]{Boroson1992} while 
${\cal R}_{\rm Fe}$ is a proxy of accretion rates \citep[e.g.,][]{Boroson1992,Marziani2003,Hu2012,Shen2014} correlating with
$Z$. 

Finally, self-gravity has been neglected in vertical direction of the BLR for its 
height (i.e., in the equation of $H=a_{0}\Omega^{-1}$). A 
sophisticated treatment of the vertical structure will include self-gravity as
well as radiation pressure from 
viscosity dissipation and star formation (and supernovae explosion). We leave this
in a future paper.

\section{Conclusions}
There is growing evidence for appearance of spiral arms in broad-line regions 
of active galactic nuclei.
In this paper, using the WKB approximation, we start from the perturbation
equations to study dynamics and structures of ionized gas in {\ccyan the SG} 
regions around SMBHs, which constitute the major parts of BLRs. 
We calculate the major properties of density waves excited by the $m=1$ mode.
The features of density waves can be detected by asymmetric profiles,
differential interferometric signals (by GRAVITY/GRAVITY+ onboard VLTI), and 2D
transfer functions from RM campaigns. In particular, the patterns of spiral arm
in the 2D transfer functions are unique due to rotation motion of spiral arm.

These features help to better understand the physical connection between SG
disks and BLRs in AGNs. Using the hypothesis that SG regions maintain the Toomre
constant $Q\sim 1$, we show that the excited density waves arising from
perturbations of SG disks are responsible for inhomogeneous distributions of
BLRs. It is possible to observationally test density waves in BLRs with the
current instruments. Our preliminary results show that density waves in BLRs
provide a new avenue for studying BLR structures and dynamics as well as for
resolving the long-standing issues of BLRs.

\vglue 0.5cm
\begin{acknowledgements}
We are grateful to an anonymous referee for a large number of comments and
suggestions from a helpful report to improve the manuscript. 
We acknowledge the support by National Key R\&D Program of China (grants 2016YFA0400701), 
by NSFC through grants {NSFC-11991054, -11991051, -12022301, -11873048, -11833008, 
-11573026}, 
and by Grant No. QYZDJ-SSW-SLH007 from the Key Research Program
of Frontier Sciences, CAS, by the Strategic Priority Research Program of the Chinese Academy of 
Sciences grant No.XDB23010400.
\end{acknowledgements}


\begin{appendix}

    \section{Basic equations}
    \label{app:basic_equations}
    We start from the ideal fluid equations in the cylindric coordinates ($R,\varphi,z$). 
    For reader's convenience, we list the classical equations which can be found from \cite{Lau1978},
    \cite{Lin1979} and \cite{Binney2008}. The continuity equation reads 
    \begin{equation}\label{eq:continuity}
    \frac{\partial\sigma}{\partial t}+
    \frac{1}{R}\frac{\partial}{\partial R}(R\sigma u)+
    \frac{1}{R}\frac{\partial}{\partial\varphi}(\sigma \vup)=0,
    \end{equation}
    and the motion euqtions are
    \begin{equation}
    \frac{\partial u}{\partial t}+
    u\frac{\partial u}{\partial R}+
    \frac{\vup}{R}\frac{\partial u}{\partial \varphi}-\frac{\vup^{2}}{R}
    =-\frac{\partial}{\partial R}(\mathscr{V}_{0}+\psi+h),
    \end{equation}
    and
    \begin{equation}
    \frac{\partial \vup}{\partial t}+
    u\frac{\partial \vup}{\partial R}+
    \frac{\vup}{R}\frac{\partial \vup}{\partial \varphi}+\frac{u\vup}{R}
    =-\frac{1}{R}\frac{\partial}{\partial \varphi}(\psi+h),
    \end{equation}
    where $\mathscr{V}_{0}$ is the potential. We should mention that the viscosity is
    neglected in above equations. This is valid for the quasi-Keplerian rotation disk
    as in the standard disk model \citep{Shakura1973}.  
    
    \section{Coefficients}
    \label{app:coefficients}
    The coefficients are given by follows
    \begin{equation}
    \calA=-\frac{1}{R}\frac{d\ln{\mathscr{A}}}{d\ln R},\quad
    {\mathscr{A}}=\frac{\kappa^{2}(1-\nu^{2})}{\sigma_{0}R},\quad
    \nu=\frac{\omega-m\Omega}{\kappa};
    \end{equation}
    \begin{equation}
    \calB=-\frac{m^{2}}{R^{2}}-\frac{4m\Omega(R\nu^{\prime})}{\kappa R^{2}\left(1-\nu^{2}\right)}+
      \frac{2m\Omega}{R^{2}\kappa\nu}\frac{d\ln\left(\kappa^2/\sigma_{0}\Omega\right)}{d\ln R},
      \,\,
    \calC=-\frac{\kappa^{2}\left(1-\nu^{2}\right)}{a_{0}^{2}}.
    \end{equation}

    \section{Wave numbers of spiral arms}
    \label{app:wavenumbers}
    
\begin{figure*}
      \centering
      \includegraphics[width= 0.95\textwidth]{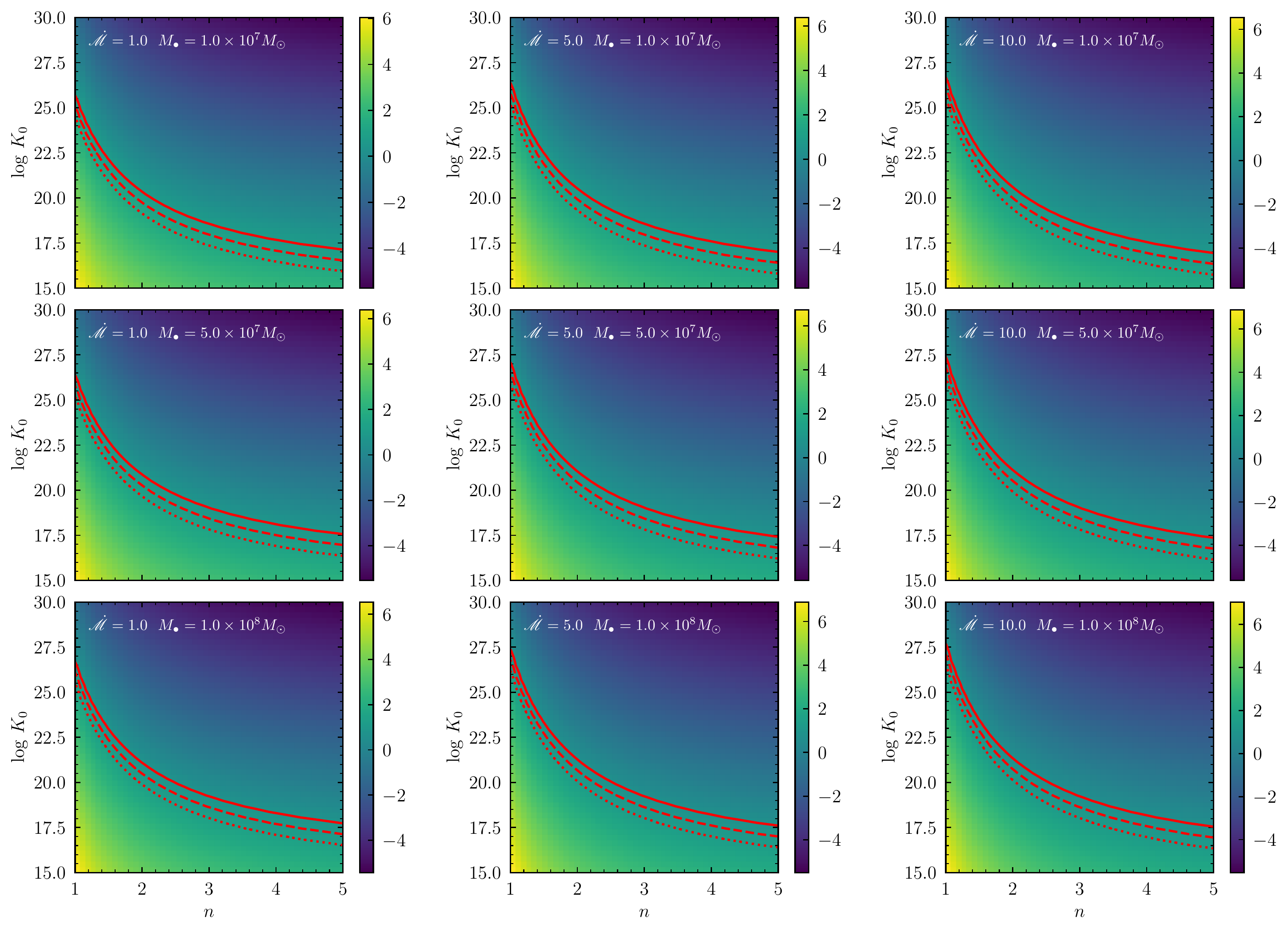}
      \caption{$\overline{k_0 R}$ as a function of $K_0$ and $n$. The
      color represents the logrithmic value of $\overline{k_0 R}$.  
      The red solid, dashed, and dotted lines represent $\overline{k_0 R} = 5, 10,
      20$, respectively (corresponding to the cases in Figure \ref{fig:spiral-arm}).}
    \label{fig:k0r}
    \end{figure*}

    We present the dependence of $\overline{k_0 R}$ on the parameter $K_0$ and
    the polytropic index $n$ in Figure \ref{fig:k0r}. $\overline{k_0 R}$
    decreases with $K_0$ and $n$ increase. In addition, $\overline{k_0 R}$
    increases slightly if $\mathdotM$ and $\bhm$ increase. The current
    dependence on $\mathdotM$ and $\bhm$ in the present model is mainly caused
    by the dependence of the inner and outer boundaries on these two parameters
    (see Eqn \ref{eq:Rtorus}). In reality, $K_0$, $n$, and even $Q_{\rm disk}$ may 
    rely on $\mathdotM$ and $\bhm$ more directly.

    \section{An example of the transfer function for unperturbed disk}
    \label{app:transfer_function_unperturbed}
    
\begin{figure}
      \centering
      \includegraphics[width=0.5\textwidth]{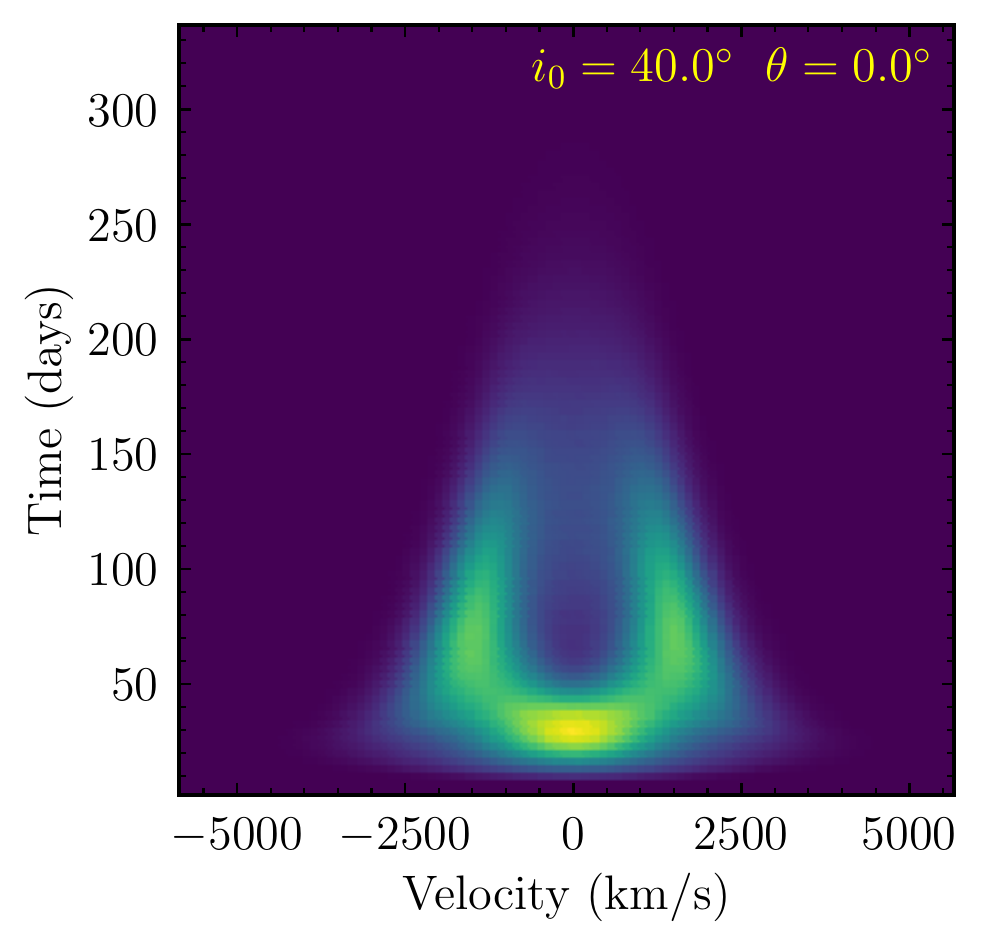}
      \caption{An example of the transfer function for unperturbed disk ($\sigma_0$).}
    \label{fig:vdm_unperturbed}
    \end{figure}

    In Figure \ref{fig:BLR} of Section \ref{sec:results}, we present the transfer
    functions of the spiral arms ($\sigma_1$). Here, for comparison, we provide an
    example of the transfer function of the unperturbed disk ($\sigma_0$) in 
    Figure \ref{fig:vdm_unperturbed}.

\end{appendix}

\end{document}